\documentclass[letter]{aa}

\usepackage{natbib}
\bibpunct{(}{)}{;}{a}{}{,}

\usepackage{graphicx}
\usepackage{epstopdf}

\usepackage{txfonts}

\newcommand{\mpl}{M_{\mathrm p}}
\newcommand{\rp}{r_{\mathrm p}}

\newcommand{\op}{\Omega_\mathrm{p}}
\newcommand{\okep}{\Omega_\mathrm{K}}
\newcommand{\cs}{c_\mathrm{s}}
\newcommand{\me}{\mathrm{M_\oplus}}
\newcommand{\msun}{\mathrm{M_\odot}}
\newcommand{\mj}{\mathrm{M_\mathrm{J}}}

\newcommand{\sigp}{\Sigma_\mathrm{p}}

\begin{document}

\title{Disc-planet interactions in subkeplerian discs}

\author{S.-J. Paardekooper}

\institute{
Department of Applied Mathematics and Theoretical Physics, University of Cambridge, Centre for Mathematical Sciences, Wilberforce Road, Cambridge CB3 0WA, United Kingdom\\
\email{S.Paardekooper@damtp.cam.ac.uk}
}

\date{Draft version \today}

\abstract{One class of protoplanetary disc models, the X-wind model, predicts strongly subkeplerian orbital gas velocities, a configuration that can be sustained by magnetic tension.}{We investigate disc-planet interactions in these subkeplerian discs, focusing on orbital migration for low-mass planets and gap formation for high-mass planets.}{We use linear calculations and nonlinear hydrodynamical simulations to measure the torque and look at gap formation. In both cases, the subkeplerian nature of the disc is treated as a fixed external constraint.}{We show that, depending on the degree to which the disc is subkeplerian, the torque on low-mass planets varies between the usual Type I torque and the one-sided outer Lindblad torque, which is also negative but an order of magnitude stronger. In strongly subkeplerian discs, corotation effects can be ignored, making migration fast and inward. Gap formation near the planet's orbit is more difficult in such discs, since there are no resonances close to the planet accommodating angular momentum transport. The location of the gap is shifted inwards with respect to the planet, leaving the planet on the outside of a surface density depression.}{Depending on the degree to which a protoplanetary disc is subkeplerian, disc-planet interactions can be very different from the usual Keplerian picture, making these discs in general more hazardous for young planets.} 
 

\maketitle
%
%________________________________________________________________

\section{Introduction}
Planets form in circumstellar discs, and the gravitational interaction between planet and disc plays a major role in shaping planetary systems. Tidal waves excited by the planet lead to orbital migration \citep{gt79}, which comes in several flavours. Type I migration \citep{ward97} is thought to hold for low-mass planets, up to a few times the mass of the Earth ($\me$). High-mass planets, comparable to Jupiter (with mass $\mj$), can tidally truncate the disc and open up a deep annular gap around their orbit. The resulting Type II migration is driven by the viscous evolution of the disc \citep{linpap86III}. A strongly dynamical form of migration, Type III \citep{maspap03}, can be achieved for intermediate-mass planets that partially open up a gap in massive discs, or by planets placed on a strong density gradient \citep{adamin}. For a recent overview of disc-planet interactions, see \cite{pap07}.  

Most studies of disc-planet interactions have focused on discs that have a Keplerian velocity profile, possibly with a slight correction for a radial pressure gradient. One class of protoplanetary disc models, however, predicts strongly subkeplerian motion due to magnetic tension \citep{shu07}. In an interesting study, \cite{adams09} point out that in these X-wind discs, migration due to the strong head wind experienced by the planet, which is moving with Keplerian velocity, can overcome Type I migration for Earth-sized planets. On the other hand, gap formation should not be affected that much, with the planet located slightly towards the inner edge of the gap.

In this Letter, we study disc-planet interactions in subkeplerian discs, under the simplifying assumption that the subkeplerian nature of the disc is an external and fixed constraint. This allows us to use purely hydrodynamical models as a first approximation. Future models should self-consistently include the evolution of the magnetic field. We outline the numerical and physical set-up in Sect. \ref{secNum}, present the results in Sect. \ref{secRes} and conclude in Sect. \ref{secCon}.
 
\section{Disc model}
\label{secNum}
The basic equations are conservation of mass, momentum, and energy in a 3D cylindrical geometry $(r,\varphi,z)$. We make two simplifications at this point: by considering an isothermal equation of state, we remove the need to solve the energy equation, and by integrating the resulting system vertically, we end up with a two-dimensional problem. Although it has been pointed out that the isothermal approximation is not valid in the inner regions of protoplanetary discs \citep{paard06, paard08}, we see below that in strongly subkeplerian discs these effects play essentially no role. The sound speed is given by $\cs=H\Omega_0$, where $H$ is the scale height of the disc and $\Omega_0$ the equilibrium angular velocity (see below). We varied $h=H/\rp$, where $\rp$ is the orbital radius of the planet, between $0.05$ and $0.2$. The higher values of $H$ are appropriate if the disc close to the star is `puffed up' by direct irradiation. A kinematic viscosity $\nu$ was used, with $\nu(r)$ such that the initial, constant surface density is a stationary solution with no accretion. In the text, we quote the corresponding $\alpha$-parameter at the location of the planet, $\alpha=\nu(\rp)/(H^2\op)$. The resulting set of equations can be found elsewhere \citep[see][]{drag}. The value of the surface density at $r=\rp$, $\sigp$, can be chosen arbitrarily in non-selfgravitating discs; we normalise the torque accordingly (see Sect. \ref{secMig}).

The gravitational potential entering the equations contains terms due to the central star and the direct and indirect component of the planet potential. We use a softened point mass potential for the planet, with a softening parameter $b=0.6H$ \citep[see][]{drag}. This value of $b$ is appropriate to account for 3D effects. The planet is assumed to be on a fixed circular orbit. For the central star potential we use $\Phi_*=-f^2\mathrm{G}M_*/r$, where $f$ is a dimensionless parameter governing the degree to which the disc is subkeplerian. It is easy to see that the resulting equilibrium angular velocity, neglecting pressure effects, is given by $\Omega_0=f\okep$, where $\okep$ is the Keplerian angular velocity. For $f=1$, we have a purely Keplerian disc, while for $f<1$, the disc is subkeplerian. X-wind models suggest $f\approx 0.6$, or even $f\approx 0.34$ during FU Orionis outbursts \citep{shu07}. 
 
We combine a linear code \citep{drag} with fully nonlinear hydrodynamic calculations with RODEO \citep{rodeo} to solve the resulting set of equations. The computational domain extends from $r/\rp=0.4$ to $r/\rp=2.5$, and covers the full $2\pi$ in azimuth. This domain is covered with a regular grid with 256 cells in the radial and 768 cells in the azimuthal direction. Nonreflecting boundary conditions were used. We vary the planet mass, $\mpl$, and $f$ to study different regimes of disc-planet interactions in subkeplerian discs.
 
\begin{figure} 
\resizebox{\hsize}{!}{\includegraphics{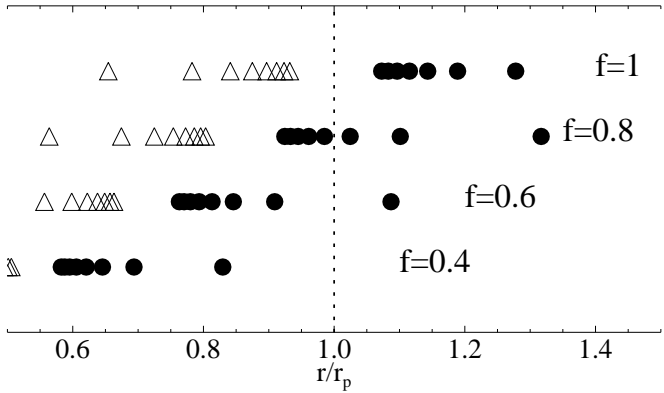}} 
\caption{Position of the inner (open triangles) and outer (solid circles) Lindblad resonances in discs with varying $f$. Pressure effects have been ignored in calculating the resonance positions.} 
\label{fig1} 
\end{figure} 
 
\section{Results}
\label{secRes}
The most important difference between subkeplerian discs and Keplerian discs, as far as disc-planet interactions are concerned, is the change in position of the Lindblad and corotation resonances. It was shown in \cite{gt79} that an embedded object can exchange angular momentum with the disc only at the location of a resonance. Corotation resonances occur where the disc corotates with the planet, and inner (outer) Lindblad resonances are located inside (outside) corotation. The strongest Lindblad resonances occur at a distance $\sim H$ of the planet \cite{ward97}. In Keplerian discs, a planet on a circular orbit is located close to corotation, with Lindblad resonances on either side of its orbit. In strongly subkeplerian discs, the distance of the planet to corotation is large enough so that we can safely neglect any nonbarotropic effects associated with the corotation torque \citep{paardpap08}. In this sense, subkeplerian discs are easier to handle, since we only have to deal with Lindblad resonances. 

The approximate position of the Lindblad resonances up to 10th order are shown in Fig. \ref{fig1} for different values of $f$. Pressure effects have been ignored in calculating these positions. In a purely Keplerian disc ($f=1$), the planet interacts with both inner and outer resonances. For $f<1$, the resonance positions move inward, with the result that the planet now mainly interacts with the outer Lindblad resonances. This has consequences for both migration and gap formation. 
 
\begin{figure} 
\resizebox{\hsize}{!}{\includegraphics{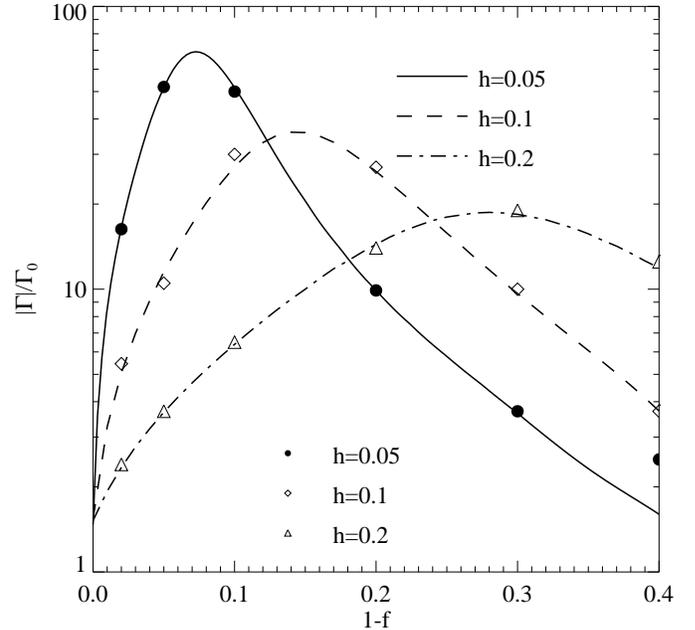}} 
\caption{Absolute value of the torque on a $q=1.26\cdot 10^{-5}$ planet embedded in discs of constant initial surface density for different values of $h$ and $f$. Solid curves denote the results of linear calculations, while the symbols denote results from hydrodynamical simulations.} 
\label{fig2} 
\end{figure} 
 
\subsection{Migration of low-mass planets}
\label{secMig}
We first consider the torque $\Gamma$ on low-mass planets, not massive enough to open up a gap. We compare results from linear calculations to hydrodynamical simulations for different values of $f$ in Fig. \ref{fig2}. The torque is normalised by $\Gamma_0=q^2\sigp\rp^4\op^2/h^2$, where $q=\mpl/M_*$ and $\op$ is the planet's angular velocity. We plot the absolute value of the torque; it is negative for all values of $f$. 

It is immediately clear from Fig. \ref{fig2} that the torque very strongly depends on $f$. This can be understood in terms of a change in the position of the resonances with respect to the planet. For $f=1$ we find the classical Type I torque, modified by nonlinear effects at corotation \citep{drag}. Around $1-f\approx h$, the planet mainly interacts with the strongest outer Lindblad resonances, yielding a very strong negative torque. The total torque approaches the one-sided Lindblad torque \citep{ward97}, which would be obtained when considering only outer Lindblad resonances and which is a factor of $\sim 1/h$ stronger than the differential Lindblad torque. The maximum torque depends on the adopted smoothing, with higher values for lower values of $b$. For $1-f \gg h$, the torque decreases, since the resonances it interacts with become weaker as the planet moves further away from corotation. For $h=0.05$ and $f=0.6$, the resulting torque is comparable in magnitude to the classical Type I torque. Linear calculations and nonlinear simulations agree very well, which again indicates that nonlinear effects at corotation do not play an important role.

The results depicted in Fig. \ref{fig2} indicate that the crucial parameter determining the total torque on the planet is $p=(1-f)/h$. The maxima of the curves occur factors of 2 apart in Fig. \ref{fig2}, confirming this picture. The scaling of the torque with $\Gamma_0$ ensures that for $f=1$, the curves fall on top of each other. When $p\approx1$, a sizeable fraction of the one-sided Lindblad torque can be exerted on the planet. The one-sided torque scales as $h^{-3}$, while the differential Lindblad torque scales as $h^{-2}$. This is the reason for the different maxima for different values of $h$. This scaling holds for all $f<1$; only for $f=1$ do we get the $h^{-2}$ scaling of the differential Lindblad torque. Therefore, in subkeplerian discs, for a fixed value of $p$, the torque scales as $h^{-3}$. For a fixed value of $f$, the situation can be different. For example, for $1-f=0.4$, the torque for $h=0.2$ is twice as strong as would be expected from an $h^{-3}$ scaling, which is due to the shift of the maximum torque to higher values of $f$.    

The torque due to a headwind in a subkeplerian disc on a planet with radius $R_\mathrm{p}$ is given by \cite{adams09}:
\begin{equation}
\Gamma_X/\Gamma_0=\frac{\pi}{2}C_\mathrm{D}(1-f)^2\frac{h}{q^2}\left(\frac{R_\mathrm{p}}{\rp}\right)^2,
\label{eqhead}
\end{equation}
with $C_\mathrm{D}$ a constant of order unity \citep{weiden77}. This means that the ratio of the torque due to a headwind and the classical Type I torque is proportional to $h$, and is of order unity for an Earth mass planet located at 1 AU, in a disc with $f=0.66$ and $h=0.05$. Since most resonances are located far inward with respect to the orbit of the planet, the actual Type I torque is close to its classical value at $f=1$ (see Fig. \ref{fig2}). It is however very difficult to generalise this result due to the complex behaviour of the Type I torque when varying $h$ and $f$.
For example, while for $f=0.6$, $\Gamma_X \approx \Gamma$ at 1 AU for a planet of 1 $\me$ in a disc with $h=0.05$, the Type I torque will in fact dominate for higher values of $h$. The dimension of the parameter space is quite high ($q$, $\rp$, $h$, etc.), so that it is difficult to make more general statements.

It is important to realise that subkeplerian discs are hazardous environments for low-mass planets. Since the planet mainly interacts with outer Lindblad resonances, and corotation torques play a minor role, migration will be directed inward. The time scale for migration can be even smaller than the classical Type I time scale, in unfavourable cases more than an order of magnitude smaller. 
       
\begin{figure} 
\resizebox{\hsize}{!}{\includegraphics{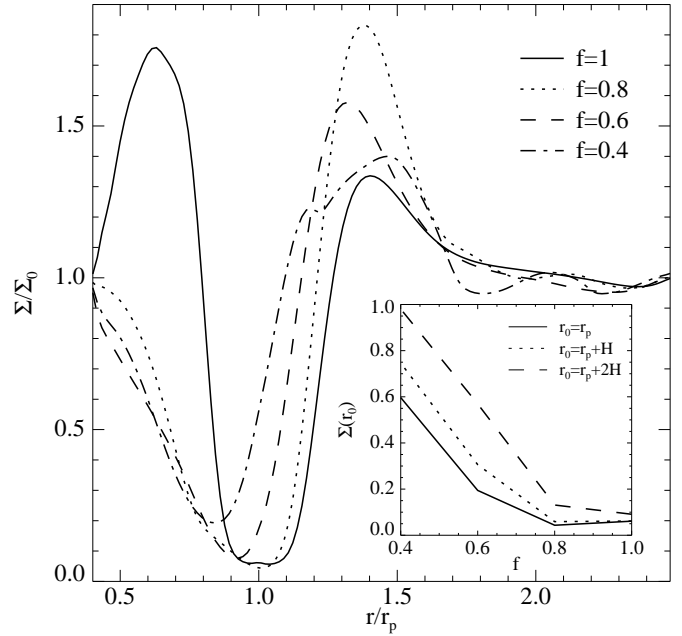}} 
\caption{Azimuthally averaged surface density, in units of the initial (constant) surface density $\Sigma_0$, for a $q=5\cdot 10^{-4}$ planet embedded in a $h=0.05$, $\alpha=0.001$ disc for different values of $f$. The inset shows the averaged surface density at the orbital radius of the planet as a function of $f$.} 
\label{fig3} 
\end{figure} 

\subsection{Gap formation}
We now turn to the issue of gap formation. The presence of a gap determines whether a planet moves in the Type I or the Type II regime, the latter being a much slower mode of migration \citep{ward97}.  There are two requirements that have to be fulfilled to open a gap: angular momentum transport due to the presence of the planet must exceed the viscous transport, otherwise the gap could be refilled by viscous evolution of the disc, and, in discs with low viscosity, the waves transporting angular momentum should be damped locally. For standard parameters, both criteria give similar minimum masses for gap opening, of the order of $100$ $\me$. A unified criterion is derived in \cite{crida06}.

\cite{adams09} derived a minimum mass for gap-opening in subkeplerian discs, $q_\mathrm{min}^{5/3} = \pi(1-f)^2\alpha h^2$. They noted that for $f=0.66$ this mass is comparable to the classical estimates for Keplerian discs, although the planet would be located more towards the inner edge of the gap. In this section, we study gap formation for a $q=5\cdot 10^{-4}$ (corresponding to $0.5$ $\mj$ around a 1 $\msun$ star) planet embedded in a disc with $h=0.05$ and a viscosity corresponding to $\alpha=0.001$. For $0.4 \leq f \leq 1$, this planet fulfils the criterion for gap formation as given above, if only slightly for $f=0.4$.

The results are depicted in Fig. \ref{fig3}, where we show the azimuthally averaged surface density for different values of $f$ after 500 orbits of the planet. The system has reached a steady state by that time. For a Keplerian disc, the surface density close to the planet has dropped by more than an order of magnitude, confirming that this planet indeed opens up a gap. The main difference between Keplerian and subkeplerian discs is the position of the gap, which is shifting inward for $f<1$. This can again be understood in terms of resonance positions: the important resonances are located inside the orbit of the planet. The associated waves are damped locally through shocks, which results in a gap at $r<\rp$.  Therefore, contrary to the findings of \cite{adams09}, the planet is located near the \emph{outside} of the gap. The analysis of \cite{adams09} only applies to the case $1-f  < h$, for which the location of the original gap (for $f=1$) is still densely populated by resonances. In this case, because the outer resonances are located closer to the planet (but still outside its orbit), the outer part of the gap forms more easily, leaving the planet closer to the inner edge of the gap.  

The depth of the gap does not strongly depend on $f$. As long as $1-f$ is not larger than the width of the gap for $f=1$, the angular momentum flux will be similar in both cases.  Then, the ability of a planet to open up a gap is relatively independent of $f$. However, from the perspective of the planet the case $f<1$ can be radically different from the Keplerian case. For what accretion and migration are concerned, what matters is the surface density close to the planet. Since the planet is located more and more outside the gap for low values of $f$, it could accrete and migrate as if there were no gap. This is illustrated in the inset of Fig. \ref{fig3}, where we show the azimuthally averaged surface density at three different locations in the disc for different values of $f$. In Keplerian discs, planets usually interact with material within 2 scale heights from their orbit \citep{bate03}. This is the maximum distance that is considered in the inset of Fig. \ref{fig3}. 

For a Keplerian disc, a gap forms with a half width that is approximately $2H$, and in the inset of Fig. \ref{fig3} all three curves fall below $0.1$ for $f=1$. For $f=0.8$, there is still a density depression around the orbit of the planet, but for lower values of $f$ there is no clear gap near $r=\rp$, it has shifted inwards enough so that the planet is basically embedded in the disc again. However, since for the lowest values of $f$ there are no more resonances located close to the planet, the Type I torque will not be fully restored. We find that typically $\Gamma(f=0.6)/\Gamma(f=1)\approx 2$ for gap-opening planets. The influence of the head wind (see Eq. \ref{eqhead}) will be strong, however, since it does not rely on resonances. We comment that a gap-opening planet is still tidally locked to the gap, just as in the Keplerian Type II migration case.

The possibility of a high-mass planet fully embedded in the disc may have some important consequences for gas accretion. In a Keplerian disc, accretion drops by an order of magnitude when a gap is formed \citep{dangelo3D}. In a subkeplerian disc, there a significant amount of mass remains near the orbit of the planet, making accretion potentially very efficient. It is not clear, however, if the planet is able to accept material that has such a high relative velocity.

\begin{figure} 
\resizebox{\hsize}{!}{\includegraphics{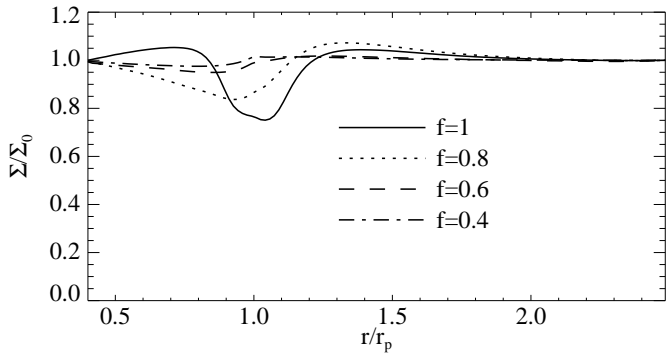}} 
\caption{Azimuthally averaged surface density, in units of the initial (constant) surface density $\Sigma_0$, for a $q=10^{-4}$ planet embedded in a $h=0.05$, $\alpha=0.004$ disc for different values of $f$.} 
\label{fig4} 
\end{figure} 

The results for a more massive planet of $q=0.001$ (1 $\mj$ around a 1 $\msun$ star) are very similar to those presented in Fig. \ref{fig3}. The gap shifts inward, and for $f<0.7$ the planet is located on the outer edge of its own density depression. For smaller planets, which only open up a shallow density depression for $f=1$, remain fully embedded for $f<1$, as shown in Fig. \ref{fig4}. While for $f=1$, a $q=10^{-4}$ planet decreases the surface density around its orbit by a factor $0.7$, for $f<0.7$ there is no more evidence for any density depression. For this lower planet mass, the important resonances become too weak to affect the surface density in strongly subkeplerian discs. 
  
\section{Discussion and conclusions}
\label{secCon}
We have presented hydrodynamical simulations of planets embedded in subkeplerian discs. They represent the first step towards modelling fully 3D magnetised discs. It is known for Keplerian discs that magnetic fields can have a strong impact on planet migration: regular fields introduce magnetic resonances \citep{terquem03}, while magnetic turbulence introduces stochastic migration \citep{nelson04,adamsbloch09}. It remains to be seen what impact a magnetic field configuration that gives rise to subkeplerian discs can have on the simple hydrodynamic picture presented here.  

We have worked in the isothermal limit, but since corotation torques only play a minor role in strongly subkeplerian discs, results for more realistic discs should be similar. The two-dimensional approximation, in combination with a gravitational softening parameter of order $h$, gives similar results to fully three-dimensional simulations, again at least as far as the Lindblad torque is concerned \citep{drag}. We have found that the results do not depend strongly on the initial surface density profile.

We have considered migration of low-mass planets (the Type I regime), finding that there is a strong dependence on $f$. For $1-f \approx h$, the planet feels almost the full one-sided Lindblad torque, which is a factor $1/h$ stronger than the classical Type I torque. Such a disc would be very hazardous to low-mass planets, since inward migration is sped up by more than an order of magnitude compared to Keplerian discs. For $1-f >h$, the torque decreases because the resonances the planet interacts with become weaker.  The dependence of the torque on $h$ and $f$ is quite complicated, and it is not easy to say for a disc of given $f$ and $h$ whether the Type I torque will be stronger of weaker than the head-wind torque. 

Gap formation proceeds similar to that in Keplerian discs. However, because of the inward shift of the important resonances, the gap will be located inside the planet's orbit. This then leaves the planet on the outside of its own gap. For strongly subkeplerian discs, a gap-opening planet can become fully embedded again. Since the torque-generating resonances are located far away, this does not restore the full Type I torque. Accretion time scales could be very short, if the planet is able to accept the available matter.

\begin{acknowledgements}
I acknowledge support from STFC in the form of a postdoctoral fellowship. I wish to thank Pawel Ivanov for his interest in retrograde orbiting embedded planets, which formed the start of this project, and the anonymous referee for an insightful report. 
\end{acknowledgements}

\bibliographystyle{aa}
\bibliography{13184.bib}

\end{document}